\documentclass[12pt,onecolumn,prb,showpacs]{revtex4}
\usepackage{graphicx}
\usepackage{amsmath}

\begin{document}

\title{Accurate freezing and melting equations for the Lennard-Jones system}

\author{Sergey A. Khrapak,$^{1,2,}\footnote{Electronic mail: skhrapak@mpe.mpg.de}$ and Gregor E. Morfill$^{1}$}

\affiliation{$^1$Max-Planck-Institut f\"ur extraterrestrische Physik, D-85741 Garching,
Germany \\$^2$Joint Institute for High Temperatures, 125412 Moscow, Russia}

\date{\today}

\begin{abstract}
Analyzing three approximate methods to locate liquid-solid coexistence in simple systems, an observation is made that all of them predict the same functional dependence of the temperature on density at freezing and melting of the conventional Lennard-Jones system. The emerging equations can be written as $T={\mathcal A}\rho^4+{\mathcal B}\rho^2$ in normalized units. We suggest to determine the values of the coefficients ${\mathcal A}$ at freezing and melting from the high-temperature limit, governed by the inverse twelfth power repulsive potential. The coefficients ${\mathcal B}$ can be determined from the triple point parameters of the LJ fluid. This produces freezing and melting equations which are exact in the high-temperature limit and at the triple point, and show remarkably good agreement with numerical simulation data in the intermediate region.
\end{abstract}

\pacs{64.60.-i, 64.70.D-}
\maketitle

Several approximate approaches have been proposed to locate liquid-solid coexistence of different substances. This includes well known phenomenological criteria for freezing and melting, like e.g.   Lindemann melting law, Hansen-Verlet freezing rule, Ravech\'{e}-Mountain-Street criterion for freezing, and a dynamical criterion for freezing in colloidal suspensions (for a review see Ref.~\onlinecite{Lowen}). These criteria are typically based on the properties of only one of the two coexisting phases and predict quasi-universal values of certain structural or dynamical quantities. Quasi-universality in this context means that a quantity is not exactly constant, but varies in a sufficiently narrow range for a broad variety of physical systems.
Although these useful empirical rules can in most cases be easily implemented with a modest computational effort, sometimes it would be desirable to have simple analytical expressions accurately describing phase coexistence, based only on the properties of the interparticle interaction potential. Even though such expressions can be non-universal (e.g. their applicability limited to a certain class of interactions), they can be quite helpful, especially in cases when interparticle interactions depend on a number of system parameters, which can vary from one situation to another.~\cite{PRL2009}

In this paper we analyze three such semi-empirical approaches developed to describe liquid-solid coexistence of simple systems and find that all of them predict the same functional dependence of the temperature on density at freezing and melting, when applied to the well known Lennard-Jones (LJ) potential of the form
\begin{equation}
U(r)=4\epsilon\left[(\sigma/r)^{12}-(\sigma/r)^{6} \right],
\end{equation}
where $r$ is the separation between a pair of particles, and $\epsilon$ and $\sigma$ are the energy and length scales. We therefore suggest to adopt this functional dependence and discuss how it can be used to produce accurate freezing and melting equations of the LJ system. The accuracy of the emerging equations is checked against existing results from numerical simulations. Some possible generalizations of this consideration are briefly discussed.

As suggested in Ref.~\onlinecite{Rosenfeld1}, for the potential $U(r)$, which is a linear combination of repulsive potentials $U_i(r)$, viz. $U(r)=\sum\alpha_iU_i(r)$, the freezing and melting curves are given, to a good approximation, by
\begin{equation}\label{additivity}
T^{{\rm L, S}}(\rho)= \sum \alpha_i T_i^{{\rm L, S}} (\rho),
\end{equation}
where $T_i^{{\rm L, S}}(\rho)$ is the temperature of the system of particles interacting via the potential $U_i(r)$, as functions of the particle density $\rho$, along liquidus (L) and solidus (S).
Equation (\ref{additivity}) describes additivity of the melting curves.~\cite{Rosenfeld1} Since the phase state of the system of particles interacting via the inverse-power-law (IPL) potential of the form $U(r)=\epsilon(\sigma/r)^n$ is determined by a single dimensionless state variable $(\rho\sigma^3)(\epsilon/T)^{3/n}$, freezing and melting are specified by the value of the parameter $c_n^{{\rm L, S}}=T^{\rm L, S}\rho^{-n/3}$, where $T$ and $\rho$ have been expressed in units of $\epsilon$ and $\sigma^{-3}$, respectively.~\cite{Comment1} For the LJ system this immediately yields
$T^{{\rm L, S}}(\rho)= 4c_{12}^{{\rm L, S}}\rho^4-4c_6^{{\rm L, S}}\rho^2$.

Another approach is based on the cell model with a spherically averaged potential in the harmonic approximation. Assuming fcc lattice in the solid phase and retaining only contributions from 12 nearest neighbors to the harmonic potential, the following melting equation has been derived in Ref.~\onlinecite{Rosenfeld2}:
\begin{equation}\label{cell_mod}
T^{\rm S}= \frac{4}{3}\delta^2[x^2U'(x)]',
\end{equation}
where $x=(\sqrt{2}/\rho)^{1/3}$ and $\delta$ is the Lindemann ratio. Applied to the LJ potential, Eq. (\ref{cell_mod}) reduces to $T^{\rm S}= 16\delta^2(11\rho^4-5\rho^2)$.
The value of the parameter $\delta$ can be chosen by fitting this equation to some known point on the melting curve.

Finally, in a recent paper\cite{PRB2010} it was observed that the freezing indicator in the form of the properly normalized second derivative of the interaction potential,  ${\mathcal L}=U''(\Delta)\Delta^2/T$ remains practically constant along the freezing curve of the LJ fluid. Here $\Delta$ is the mean interparticle distance (related to the non-reduced density via $\Delta=\rho^{-1/3}$) and $T$ is non-reduced temperature.
The emerging equation for freezing of the LJ fluid can be written in reduced units as $T^{\rm L}=(24/{\mathcal L})(26\rho^4-7\rho^2)$ The value of the constant ${\mathcal L}$ can be again determined by fitting this equation to some freezing point of the LJ fluid. For example, the high-temperature limit, governed by $r^{-12}$ repulsion, can be used. In this way the value ${\mathcal L}\simeq 290$ has been estimated.~\cite{PRB2010}

All the considered approaches are approximate in their nature  and involve serious simplifications . However, the very fact that three different (although not completely independent considerations)
lead to the same functional form of the liquid-solid coexistence boundaries is indicative. It strongly suggests to use the functional form
\begin{equation}\label{ff}
T^{{\rm L, S}}={\mathcal A}^{{\rm L, S}}\rho^4-{\mathcal B}^{{\rm L, S}}\rho^2
\end{equation}
to describe freezing and melting curves of the LJ system. Our main purpose here is not to discuss which of the approaches discussed above provides better values of the coefficients ${\mathcal A}$ and ${\mathcal B}$. Instead, we suggest a procedure to determine the coefficients in Eq.~(\ref{ff}), which results in highly accurate freezing and melting equations of the LJ system applicable in the entire temperature range, from the triple point up to the high temperature limit.

The coefficients ${\mathcal A}$ and ${\mathcal B}$ can be determined from the two reference points on the freezing/melting curve. One natural choice of the reference point is to take temperatures so high that the attractive contribution in the LJ potential can be ignored. In this regime the following dependence of density on temperature has been reported: $\rho^{\rm L}\simeq 0.813 T^{1/4}$ at freezing and $\rho^{\rm S}\simeq 0.844 T^{1/4}$ at melting.~\cite{Hoover1970,Hansen}   This fixes the values of the coefficients ${\mathcal A}$ at freezing and melting. We get ${\mathcal A}^{\rm L}\simeq 2.29$ and ${\mathcal A}^{\rm S}\simeq 1.97$. To determine the coefficients ${\mathcal B}$ in Eq.~(\ref{ff}) it is appropriate to look at the low-temperature region of the liquid-solid coexistence. In particular, triple point parameters of the LJ system are especially convenient for this purpose. The triple point temperature and densities obtained in different numerical simulations are summarized in Table \ref{Tab}. There is some scattering of the data points, which is apparently due to differences in simulation methods as well as computational details such as system size, finite cutoff radius, etc. We will not discuss here the accuracy of each particular simulation. Instead, we take some ``average'' values for the triple point temperature and densities: $T_{\rm tr}\simeq 0.68$, $\rho_{\rm tr}^{\rm L}\simeq 0.85$ and $\rho_{\rm tr}^{\rm S}\simeq 0.96$. These are also the values suggested by Hansen and McDonald.~\cite{Hansen_Book} This estimate is accurate to within few percent and can be further improved when consensus regarding the exact triple point parameters of the LJ system is reached. The emerging values of the coefficients ${\mathcal B}$ at freezing and melting are ${\mathcal B}^{\rm L}\simeq 0.71$ and ${\mathcal B}^{\rm S}\simeq 1.08$, respectively.

\begin{table}
\caption{\label{Tab} Triple point parameters of the LJ fluid obtained in different numerical simulations.}
\begin{ruledtabular}
\begin{tabular}{llll}
$T_{\rm tr}$ & $\rho_{\rm tr}^{\rm L}$ & $\rho_{\rm tr}^{\rm S}$ & Ref.  \\ \hline
0.660 & 0.862 & 0.960 & [\onlinecite{Hansen_1969}]  \\
0.687 & 0.850 & 0.963 & [\onlinecite{Agrawal}]  \\
0.692 & 0.847 & 0.962 & [\onlinecite{Barroso}]  \\
0.661 & 0.864 & 0.978 & [\onlinecite{Ahmed}]  \\
\end{tabular}
\end{ruledtabular}
\end{table}

\begin{figure}
\includegraphics[width=8.2cm]{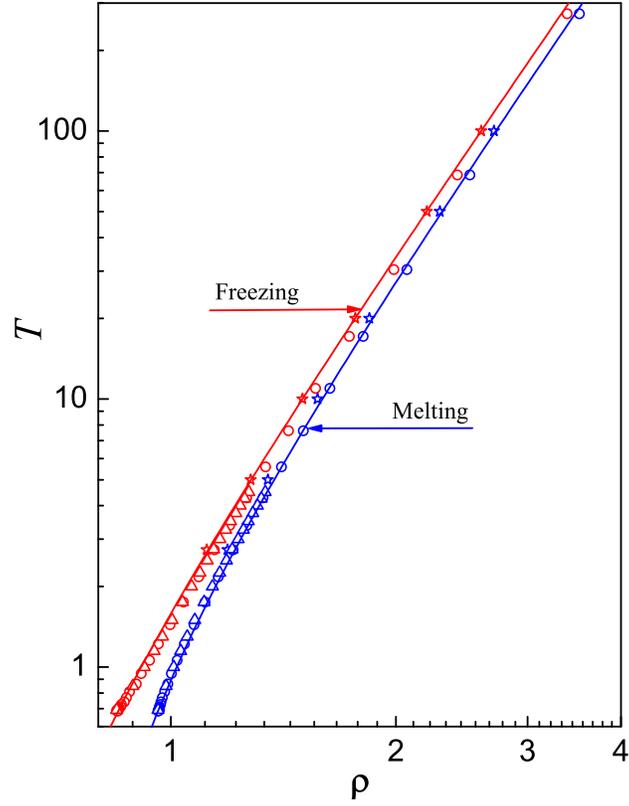}
\caption{Comparison of the freezing and melting equations proposed in this paper (solid curves) with the representative set of numerical data related to liquid-solid coexistence of the Lennard-Jones system (symbols). For each value of the reduced temperature there are two values of reduced density. The left (right) symbol corresponds to freezing (melting). Stars, crosses, and triangles denote numerical results from Refs.~\onlinecite{Hansen}, \onlinecite{Agrawal}, and \onlinecite{Barroso}, respectively.} \label{f1}
\end{figure}

Expression (\ref{ff}) with the coefficients ${\mathcal A}$ and ${\mathcal B}$ derived above constitute freezing and melting equations of the LJ system, which are ``exact'' in the high-density high-temperature limit and at the triple point. In the intermediate regime, agreement with numerical simulation data is also remarkable, as shown in Fig.~\ref{f1}. The derived equations represent a considerable improvement over phenomenological fits proposed in the literature,~\cite{Ahmed,Mastny,vanHoef} which are sufficiently accurate only in the vicinity of the triple point.

It is important to point out that similar freezing and melting equations can be constructed for a wide range of the LJ-type potentials. Obvious example is related to $n$-6 and more general Mie ($n$,$m$) model potentials. However, we do not elaborate further on this possibility, because triple point parameters of these systems are known with much lower accuracy.~\cite{Ahmed,KhrapakJCP}

To summarize, we propose very simple analytic equations for freezing and melting of the LJ system, which are very accurate in the entire range of temperatures and densities. These equations are superior to phenomenological fits proposed earlier. Similar equations can be easily derived for other model potentials of the LJ type.

This work was partly supported by DLR under Grant 50WP0203.
(Gef\"{o}rdert von der Raumfahrt-Agentur des Deutschen Zentrums f\"{u}r Luft und Raumfahrt e. V. mit Mitteln des Bundesministeriums f\"{u}r Wirtschaft und Technologie aufgrund eines Beschlusses des Deutschen Bundestages unter dem F\"{o}rderkennzeichen 50 WP 0203.)

\begin{thebibliography}{99}
\bibitem{Comment1} This normalization is used throughout the paper, except when specially indicated.
\end{thebibliography}
\end{document}